\documentclass[]{agujournal2019}
\usepackage{url} 
\usepackage[inline]{trackchanges} 
\usepackage{soul}
\usepackage{bbding}
\usepackage{graphicx}
\usepackage{float}
\usepackage{amssymb, amsmath}
\graphicspath{{Figs/}}

\usepackage{todonotes}

\draftfalse

\journalname{Space Weather}
\draftfalse

\begin{document}

%
%

\title{Multi-Hour Ahead Dst Index Prediction Using Multi-Fidelity Boosted Neural Networks}

%
%




\authors{A. Hu\affil{1}, E. Camporeale\affil{1,2}, B. Swiger\affil{1,2}}

\affiliation{1}{CIRES, University of Colorado, Boulder, CO, USA}
\affiliation{2}{NOAA Space Weather Prediction Center, Boulder, CO, USA}




\correspondingauthor{A. Hu}{andong.hu@colorado.edu}




\begin{keypoints}
\item A new multi-hour ahead $Dst$ prediction model developed from solar wind observations using Gated Recurrent Unit (GRU) networks is proposed.
\item The uncertainty of the proposed $Dst$ model is estimated by applying the ACCRUE method.
\item A multi-fidelity method is developed to boost the performance of the model.

\end{keypoints}

%
%

%
%


\begin{abstract}

The Disturbance storm time ($Dst$) index has been widely used as a proxy for the ring current intensity, and therefore as a measure of geomagnetic activity. It is derived by measurements from four ground magnetometers in the geomagnetic equatorial regions. 
We present a new model for predicting $Dst$  with a lead time between 1 and 6 hours. The model is first developed using a Gated Recurrent Unit (GRU) network that is trained using solar wind parameters. The uncertainty of the $Dst$ model is then estimated by using the ACCRUE method [Camporeale et al. 2021]. Finally, a multi-fidelity boosting method is developed in order to enhance the accuracy of the model and reduce its associated uncertainty. It is shown that the developed model can predict $Dst$ 6 hours ahead with a root-mean-square-error (RMSE) of 13.54 $\mathrm{nT}$. This is significantly better than the persistence model and a simple GRU model. 
\end{abstract}

\section*{Plain Language Summary}

Geomagnetic storms pose one of the most severe space weather risks to our space borne and ground-based electronic instruments, such as GNSS and radio transmission systems. Dst is one of the most accurate geomagnetic storm indicators. Hence, those storm can be predictable if Dst can be predicted. This study presents an innovative multi-fidelity boosted neural network method to improve the performance of the 1-to-6 hours predictions model by considering the uncertainty of the predictions.

%
%

%


%
%
%
%

\section{Introduction}
\label{sec:introduction}

The Disturbance storm time ($Dst$) is a geomagnetic index related to the perturbation of the geomagnetic field at low latitudes \cite{burton1975empirical, rostoker1972geomagnetic}. Currently, $Dst$ is defined by using geomagnetic field measurements from four equatorial ground magnetometers: Hermanus, Honolulu, San Juan and Kakioka \cite{sugiura1991equatorial}. $Dst$ has been widely used for monitoring geomagnetic storms which pose one of the most severe space weather risks to our space-borne and ground-based electronic instruments, such as GNSS and radio transmission systems \cite{wan2021nighttime, li2021status}.

A short-term prediction of Dst is produced operationally at the NOAA Space Weather Prediction Center (SWPC) by means of a physics-based model (the Space Weather Modeling Framework developed at the University of Michigan, \cite{toth2005space}).
A longer lead-time operational Dst forecast is provided by Space Environment Technologies, using the Anemomilos model \cite{tobiska2013anemomilos}.
$Dst$ is also used as an essential input for forecasting thermosphere mass density and ionospheric parameters, and to parameterize several empirical models. A non-exhaustive list of
models that use Dst as one of their inputs includes: \citeA{o2003empirical} empirically
estimate the location of the plasmapause; \citeA{agapitov2015empirical} derive a statistical model for the lower band chorus distribution; \citeA{li2016empirical} estimate the ionospheric global electron content storm-time response; \citeA{boardsen2000empirical} derive an empirical model of the high‐latitude magnetopause; \citeA{zhao2018empirical} derive a model of radiation belt electron pitch-angle distributions.

A large amount of literature has been devoted to Dst prediction, notably using empirical and machine learning techniques \cite{camporeale2019challenge}. 
\citeA{lundstedt2002operational} first attempted to implement a multi-layer perception (MLP) neural network to forecast $Dst$ one hour ahead using interplanetary magnetic field (IMF) data. Several researchers presented models to extend $Dst$ forecast up to 6-hrs in advance \cite{saiz2008forecasting, bala2012improvements, lazzus2017forecasting}. A Gaussian Process model was introduced by \citeA{chandorkar2017probabilistic} and combined with a long short-term memory (LSTM) architecture in \citeA{gruet2018multiple} to provide probabilistic predictions up to six hours in advance. An ensemble learning algorithm was applied by \citeA{xu2020prediction} for the same purpose.
\citeA{laperre2020dynamic} evaluated the performance of a LSTM model based on a Dynamical Time Warping (DTW) metric as the cost function. 

In this study, we first train a model using a machine learning (ML) technique called Gated Recurrent Unit (GRU), which is a flavor of a recurrent neural network (RNN, \citeA{hu2018using}), to forecast $Dst$ during strong storm periods \add[]{($Dst\,<\,-100 \mathrm{nT}$)} 1-6 hours ahead. 
The corresponding uncertainties associated to the predictions, that we refer to as $\Delta Dst$, are then estimated by using the ACCRUE method \cite{camporeale2021accrue}, also using a GRU network.   
{The multi-fidelity boosting method proposed here works as follows. The accurate estimate of the uncertainty $\Delta Dst$ for a given trained model, can inform us about the input conditions under which the model does not perform well. Hence, we can identify a subset of the original training set that can be used for training a different, independent, model. Such a strategy can be iterated a number of times. The final result will be a collection of models, each working very well for a specific subset of input conditions (hence the built-in multi-fidelity). The crucial point, though, is that since each $Dst$ model comes with its own estimate of uncertainty, that can be used as a weighting factor when optimally combining (in a linear fashion) a large number of models. }

The paper is divided as follows. Section \ref{sec:D&M} introduces the data used for this study, the criteria to define storm events and the corresponding time periods covered. The methodology, including the designed UQ-based machine learning architecture and developed multi-fidelity boosting method, is also described. Section \ref{sec:results} presents the results of the developed model, and discusses the advantages of the proposed model. Finally, in Section \ref{sec:summary-outlook}, we draw conclusions and make final remarks about future directions.

\section{Data and Methods}
\label{sec:D&M}

\subsection{Data}
\label{subsec:Data}

Studies done in the past to predict the geomagnetic index Dst have shown that various solar wind parameters are of interest to optimize the performance of predicting models. \citeA{gruet2018multiple} selected the electron density $n$, the solar wind velocity $V$, IMF $|B|$ and $B_z$. The same variables are considered in this study. In addition to that, several variables used to estimate the geomagnetic field as in \citeA{weimer2013empirical} and the historical value of geomagnetic field perturbations from SuperMAG are taken into account. All variables are shown in Table \ref{tab:variables}. It should be noted that one-day-before $F10.7$ is selected instead of the real-time $F10.7$ because $F10.7$ is a daily average which is unavailable in real time. 
The models will be trained by using this variable set defined between $t-delay-6$ and $t-delay$, where $t$ is the predict time stamp and $delay$ is how many hours ahead we are going to predict.

\begin{table}[!htbp]
\centering
\caption{List of variables.}
\begin{tabular}{c c c}
  \hline
  Reference & Variable & Symbol\\
  \hline
  & electron density & $n$ \\  
  & solar wind velocity & $V$ \\
  \citeA{gruet2018multiple} & Norm of IMF magnetic field vector & $|B|$ \\
  & IMF magnetic field vector in z direction & $B_z$ \\ 
  & $Dst$ & $Dst$ \\
  \hline
  & clock angle from tangential IMF in the GSM Y-Z plane & $\theta_c$ \\
  \citeA{weimer2013empirical} & dipole tilt angle & $dipole$ \\
  & square root of F10.7 one day  before & $\sqrt{F10.7}$  \\
\end{tabular}
\label{tab:variables}
\end{table}

\subsubsection{Disturbance storm time ($Dst$) index}
\label{subsubsec:Dst}

\begin{figure}
  \centering
  \includegraphics[width=\textwidth]{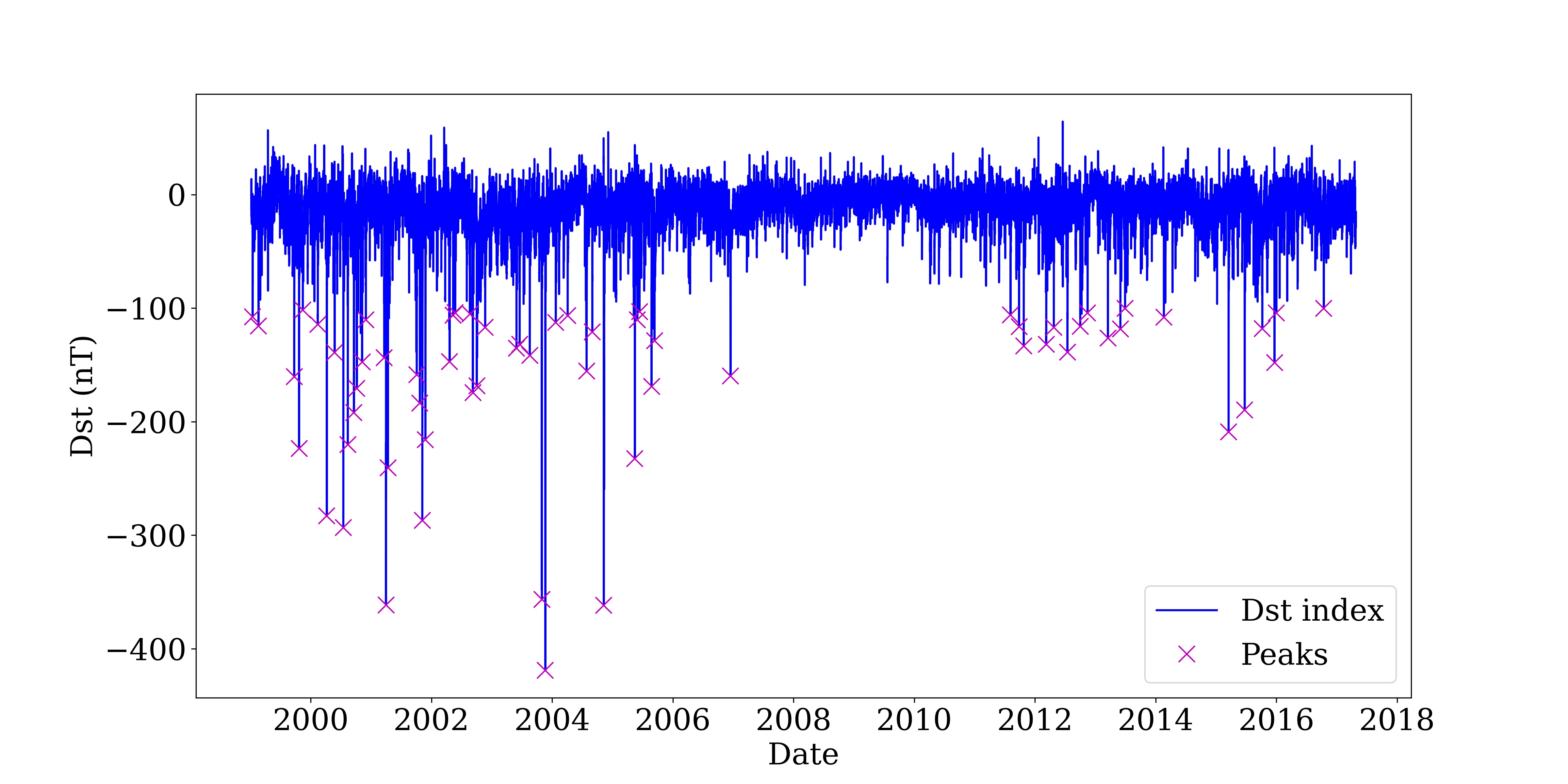}
  \caption{Time history of $Dst$ during the period 1999 to 2017. Time on horizontal axis and $Dst$ values on vertical axis. The magenta crosses denote peak values smaller than -100 nT, used for defining storm events considered in this study.}
  \label{fig:Peak_select}
\end{figure}

The historical $Dst$ index is available at 1-hour cadence from the NASA OMNI database. Fig. \ref{fig:Peak_select} displays the $Dst$ index in the period 1999-2017 . The model is trained and tested on storm events with a $Dst$ peak smaller than -100 nT, shown by magenta crosses in Fig. \ref{fig:Peak_select}. Overall, 67 such storm periods are selected for this study. 

\begin{figure}
  \centering
  \includegraphics[width=0.7\textwidth]{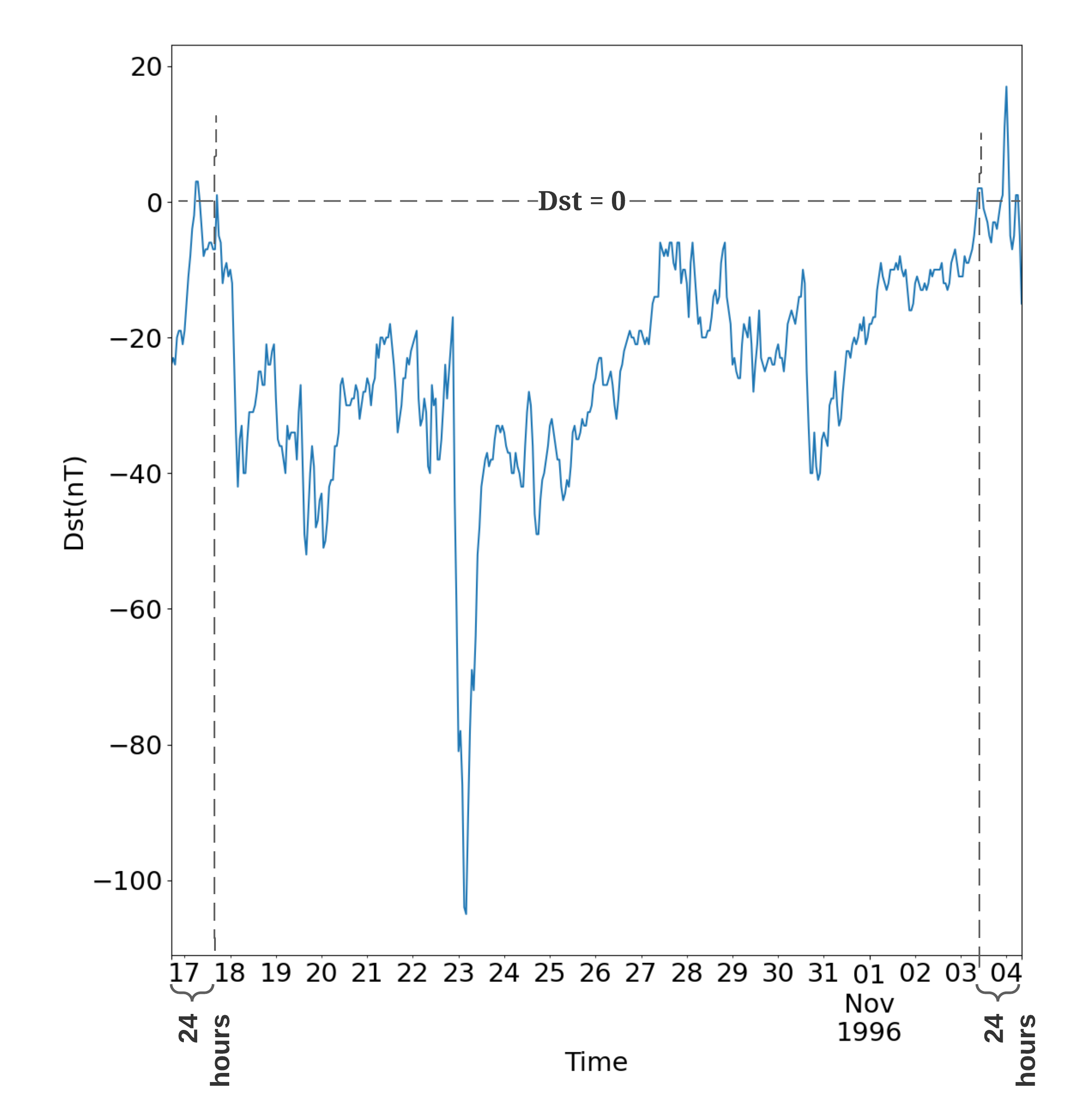}
  \caption{An example of the selection criterion used to define the time range for one storm event. The $Dst$ peak occurs on Oct. 23, 1996. The nearest positive $Dst$ values before and after the peak occur on Oct. 18 and Nov. 03, respectively. The whole storm range is defined between Oct. 17, 1996 and Nov. 04, 1996 with a 24-hour buffer zone. The list of selected storm events can be found in Table. \ref{tab:storms}.}
  \label{fig:range_select}
\end{figure}

Consistently defining a storm period is a difficult task \cite{gonzalez1994geomagnetic}, since it usually includes a pre-storm period, a main phase and a recovery phase.  In this study, we define a storm event by looking for the nearest positive $Dst$ values immediately before and after each negative peak, and then extending the time window by a 24-hour buffer. An example of such a definition of a storm event is shown in Fig. \ref{fig:range_select}, where the Dst peak is observed on Oct. 23, 1996. The storm period is defined as ranging between Oct. 17, 1996 and Nov. 04, 1996. With this procedure we make sure that the time intervals are selected in such a way that the negative $Dst$ peaks do not always occur at the same time within the chosen storm-time window. Hence, the neural network does not memorize. The average period of selected storm events is approximately 15 days. All 67 selected storms are listed in Table \ref{tab:storms}.

\begin{table}[!htbp]
\centering
\caption{List of first 67 Storm Events.}
\begin{tabular}{c c c | c c c}
  \hline
  Start date/time & End date/time & Min. Dst (nT) &
  Start date/time & End date/time & Min. Dst (nT)\\
  \hline

2000-02-10T01 & 2000-02-19T06 & -135.0 & 2004-01-20T23 & 2004-01-29T23 & -130.0\\
2000-03-31T05 & 2000-04-12T13 & -292.0 & 2004-04-02T09 & 2004-04-05T00 & -117.0\\
2000-05-22T19 & 2000-06-02T09 & -147.0 & 2004-07-23T05 & 2004-08-06T12 & -170.0\\
2000-07-14T09 & 2000-07-19T12 & -300.0 & 2004-08-28T20 & 2004-09-03T19 & -129.0\\
2000-08-08T23 & 2000-08-19T01 & -234.0 & 2004-11-06T14 & 2004-11-19T11 & -374.0\\
2000-09-16T13 & 2000-09-24T08 & -201.0 & 2005-01-15T15 & 2005-01-21T11 & -103.0\\
2000-09-23T11 & 2000-10-12T16 & -181.0 & 2005-05-13T23 & 2005-05-19T22 & -247.0\\
2000-11-02T22 & 2000-11-10T00 & -159.0 & 2005-05-28T16 & 2005-06-04T05 & -113.0\\
2000-11-25T16 & 2000-12-02T23 & -119.0 & 2005-06-11T11 & 2005-06-16T00 & -106.0\\
2001-03-18T05 & 2001-03-22T08 & -149.0 & 2005-08-23T02 & 2005-08-30T21 & -184.0\\
2001-03-29T21 & 2001-04-04T09 & -387.0 & 2005-09-08T08 & 2005-09-24T21 & -139.0\\
2001-04-07T06 & 2001-04-21T10 & -271.0 & 2006-12-04T17 & 2006-12-31T20 & -162.0\\
2001-04-20T18 & 2001-04-25T08 & -102.0 & 2011-08-04T14 & 2011-08-13T15 & -115.0\\
2001-08-16T10 & 2001-08-19T06 & -105.0 & 2011-09-25T09 & 2011-09-30T17 & -118.0\\
2001-09-24T15 & 2001-10-07T02 & -166.0 & 2011-10-23T15 & 2011-10-30T04 & -147.0\\
2001-10-18T04 & 2001-10-25T03 & -187.0 & 2012-03-07T07 & 2012-03-22T11 & -145.0\\
2001-11-04T13 & 2001-11-14T18 & -292.0 & 2012-04-22T11 & 2012-05-05T06 & -120.0\\
2001-11-23T00 & 2001-11-30T08 & -221.0 & 2012-07-13T19 & 2012-07-20T07 & -139.0\\
2002-03-22T09 & 2002-03-28T15 & -100.0 & 2012-09-29T08 & 2012-10-05T02 & -122.0\\
2002-04-16T05 & 2002-04-27T01 & -149.0 & 2012-11-12T11 & 2012-11-17T03 & -108.0\\
2002-05-10T05 & 2002-05-18T14 & -110.0 & 2013-03-16T00 & 2013-03-23T17 & -132.0\\
2002-05-22T05 & 2002-05-30T14 & -109.0 & 2013-05-30T19 & 2013-06-08T23 & -124.0\\
2002-07-31T17 & 2002-08-06T21 & -102.0 & 2013-06-26T19 & 2013-07-04T12 & -102.0\\
2002-08-17T16 & 2002-08-25T04 & -106.0 & 2014-02-17T07 & 2014-02-23T01 & -119.0\\
2002-09-02T17 & 2002-09-16T01 & -181.0 & 2015-03-16T00 & 2015-03-25T06 & -223.0\\
2002-09-28T20 & 2002-10-23T09 & -176.0 & 2015-06-21T05 & 2015-07-02T22 & -204.0\\
2002-09-28T20 & 2002-10-23T09 & -100.0 & 2015-09-30T07 & 2015-10-17T14 & -124.0\\
2002-11-16T04 & 2002-12-06T10 & -128.0 & 2015-12-18T22 & 2015-12-24T07 & -155.0\\
2003-05-20T08 & 2003-06-07T14 & -144.0 & 2015-12-30T05 & 2016-01-04T05 & -110.0\\
2003-06-15T03 & 2003-06-26T06 & -141.0 & 2016-10-12T00 & 2016-10-20T01 & -104.0\\
2003-07-09T19 & 2003-07-23T01 & -105.0 & 2017-05-26T16 & 2017-05-31T13 & -125.0\\
2003-08-16T12 & 2003-08-31T17 & -148.0 & 2017-09-05T22 & 2017-09-14T06 & -142.0\\
2003-10-23T20 & 2003-11-04T00 & -383.0 & 2018-08-24T11 & 2018-09-01T05 & -174.0\\
2003-11-07T22 & 2003-11-29T14 & -422.0\\				
  
\end{tabular}
\label{tab:storms}
\end{table}

\subsection{Methodology}
\label{subsec:methods}

\subsubsection{Gated Recurrent Unit (GRU) Neural Networks}
\label{subsubsec:GRU}

\begin{figure}[!htbp]
	\centering
    \includegraphics[width=0.7\textwidth]{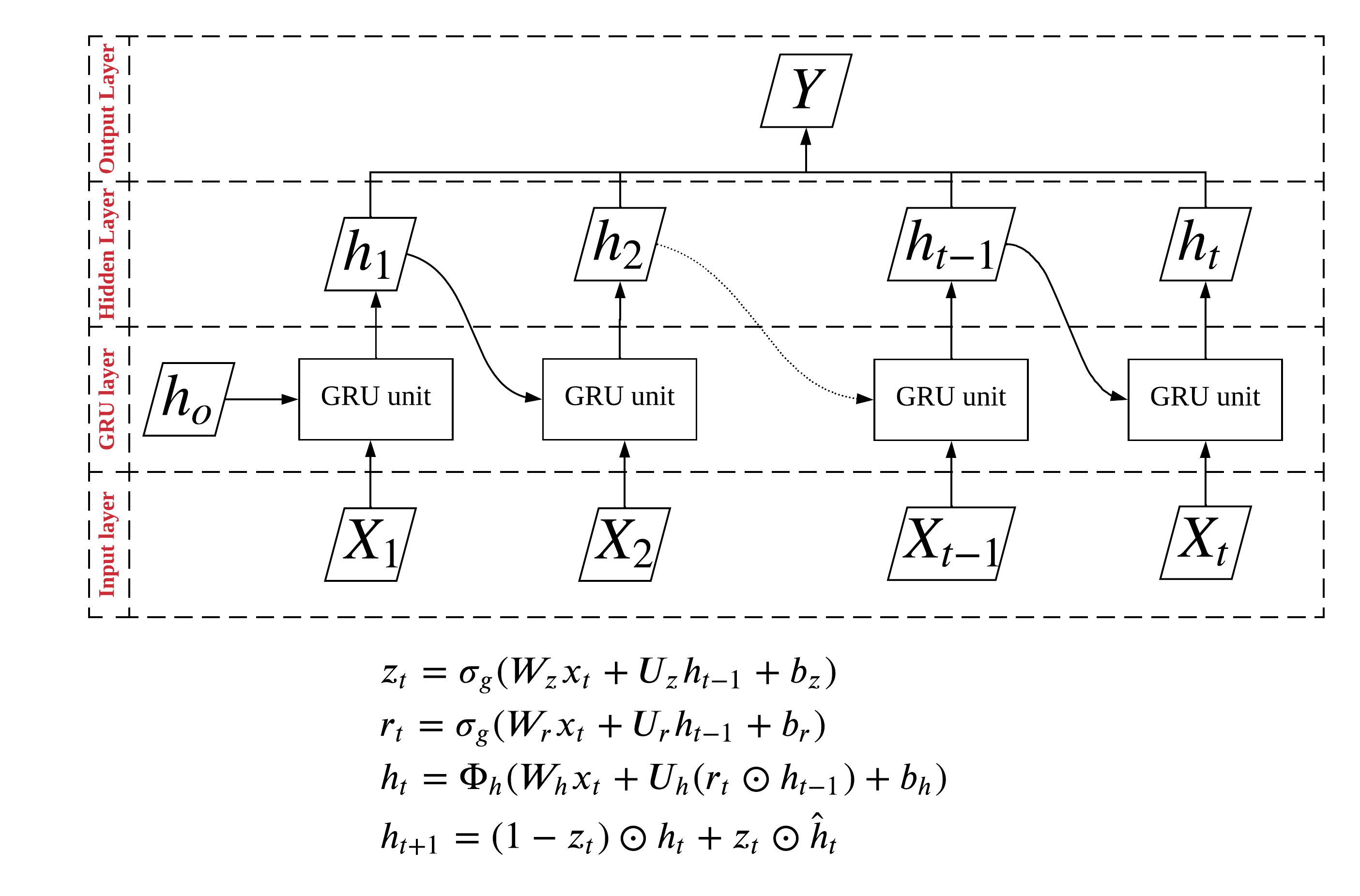}
    \caption{Structure of GRU. $x_t$ is the independent variable set at the time $t$, and $Y$ is the target. $h_t$ is the temporary results from the $t$th GRU unit, $h_0$ is manually initialized. The connection between hidden layer and output layer is by means of a simple fully-connected layer. Each GRU unit can be considered as a MLP model. $z_t$ is an update gate vector and $r_t$ is a reset gate vector. $W$, $U$ and $b$ are the coefficients that need to be estimated during training. In addition, $\sigma_g$ and $\Phi_h$ are sigmoid and tanh activation functions, respectively.}
    \label{fig:GRU}
\end{figure}

Gated Recurrent Unit (GRU) networks are one of the most widely used Recurrent Neural Networks (RNNs). Similar to Long Short-Term Memory (LSTM), GRU was proposed as a solution to short-term memory and vanishing gradient problems \cite{mikolov2010recurrent}. In most scenarios, the performance of GRU is on par with LSTM, but computationally more efficient because of a less complex structure \cite{kaiser2015neural}. The architecture of GRU is shown in Fig. \ref{fig:GRU}. $x_t$ is formed by all variables in Table \ref{tab:variables} at a certain time $t$. $X$ is a time series with a 6-hr span and a 1-hr time step. 
$Y$ is the corresponding $Dst$ with a fixed time delay, i.e., 1-6 hours.

\subsubsection{ACCRUE method}
\label{subsubsec:ACCRUE}

The ACCRUE method is proposed by \citeA{camporeale2021accrue} for assigning uncertainties to single-point predictions generated by a deterministic model that outputs a continuous variable. The problem of estimating the optimal uncertainty is set up as an optimization problem, where the ACCRUE score is defined as the combination of the continuous ranked probability score (CRPS) and the Reliability Score (RS) according to a certain ratio. In $\Delta Dst$ modeling, the ACCRUE score is used as the cost function for the same GRU architecture utilized for the corresponding $Dst$ model. The main assumption of the ACCRUE method is that the residuals of the underlying deterministic model (i.e., the difference between the model output and the ground truth) are normally distributed. The model output is taken as the mean of a Gaussian distribution, and its uncertainty is assigned by estimating the corresponding standard deviation.
Note that the ACCRUE model is trained in a self-supervised regression mode, because the target (i.e. the standard deviation) is unknown, and it enters in the minimized loss function as a non-trivial, yet analytical, function.
If the ACCRUE model is successfully trained, then the distribution of the so-called z-scores, that is the algebraic errors divided by the corresponding standard deviations, collected over the whole training set, should also be normally distributed. A graphical technique called quantile-quantile plots (Q-Q plots)\cite{wilk1968probability}, is then applied to investigate whether the ACCRUE results indeed return a normal distribution for the z-scores.

\subsubsection{Multi-fidelity Boosting Method}
\label{subsub:En-boost}
The multi-fidelity boosting strategy is used to develop a collection of models trained on a subset of the original training set. For a given model, the ACCRUE method is applied in order to estimate the associated uncertainties (trained on the errors evaluated over the whole the training set). \cite{camporeale2021accrue}. All training samples are then sorted according to a chosen accuracy criterion. The best 10\% samples are discarded in subsequent iterations, while a new subset is defined by the worst 50\% samples as the next training set, used to train an independent model. Such a strategy can be iterated a number of times to form an ensemble of multi-fidelity models. Finally, by exploiting the knowledge about the uncertainties associated with each model, the final predictions, which are expected to outperform the predictions of each single model, are defined as a weighted combination of each ensemble member. Three criteria are tested in this study. They are readily defined for each sample in the training set, and they are: the absolute error, the standard deviation as estimated from ACCRUE ($\sigma$) and their z-score defined as $\frac{error}{\sigma}$. When developing a model that predicts $N$ hours ahead, we use the final model previously developed for $N-1$ hours as our baseline model, from which the iterative methods starts. The final 

\begin{figure}[htbp]
  \includegraphics[width=1\textwidth]{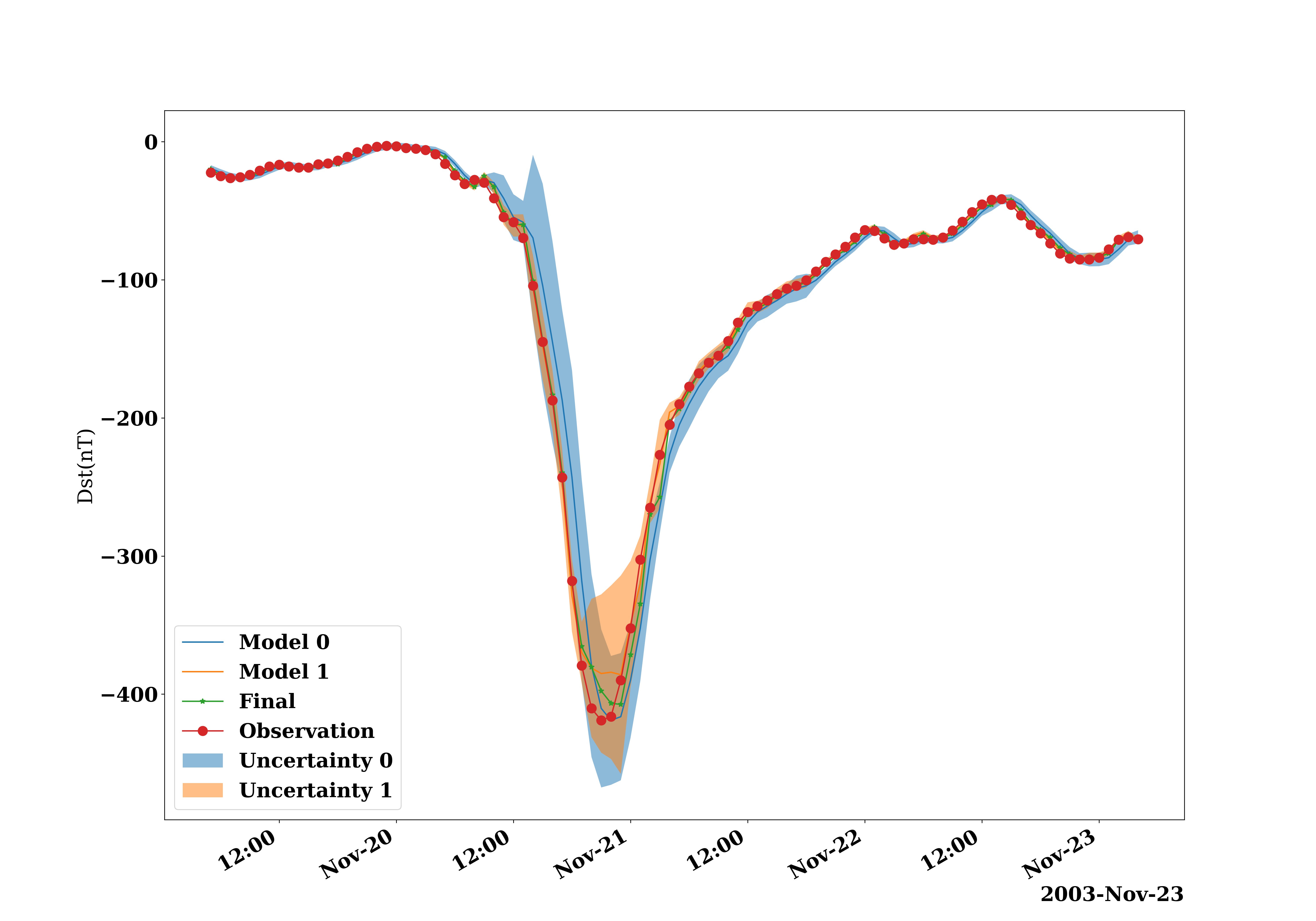}
\caption{Illustration of multi-fidelity boosting method for one storm event. Time is on horizontal axis and $Dst$ on vertical axis. Blue and orange lines are from trained models. Blue and orange shaded areas are the corresponding standard deviations. Green line is the final predictions, and red dot line is the observed Dst.}
  \label{fig:ensemble}
\end{figure}

An example of the proposed multi-fidelity boosting method for a given storm event is shown in Fig. \ref{fig:ensemble}. Here, `model\ 0' is a simple persistence model. It is clear that `model\ 1' (obtained after one iteration of the boosting method) performs better during the pre-storm and recovery phase, and `model\ 0' performs better during the main phase of the storm. The combination of these two models can outperform each if their uncertainty can be well estimated. The final predictions are shown as Eqn. \ref{eqn:final}, where $pred_f$ and $pred_i$ denote final predictions and predictions from the i$th$ boost model, and $\sigma_i$ is the uncertainty  corresponding to the i$th$ boost model.

\begin{align}
\label{eqn:final}
    pred_f = \sum_{i=1}^n\frac{\frac{1}{\sigma_i^2}}{\sum_{j=1}^n\frac{1}{\sigma_j^2}}\times pred_i
\end{align}

The accuracy of the models developed with different criteria are shown in table \ref{tab:boost}. 20\% of all samples are selected for this validation. Those samples will not be used for training. It is clear that the model developed with  `z-score' outperforms the other two significantly within all $Dst$ ranges. Hence,  `z-score' is selected for the following study. All training samples will be separated into three subsets for training, early stop, and the aforementioned criteria validation. The training set needs to be shuffled before training in order to make sure the training set is homogeneously distributed.

\begin{figure}[htbp]
  \includegraphics[width=1\textwidth]{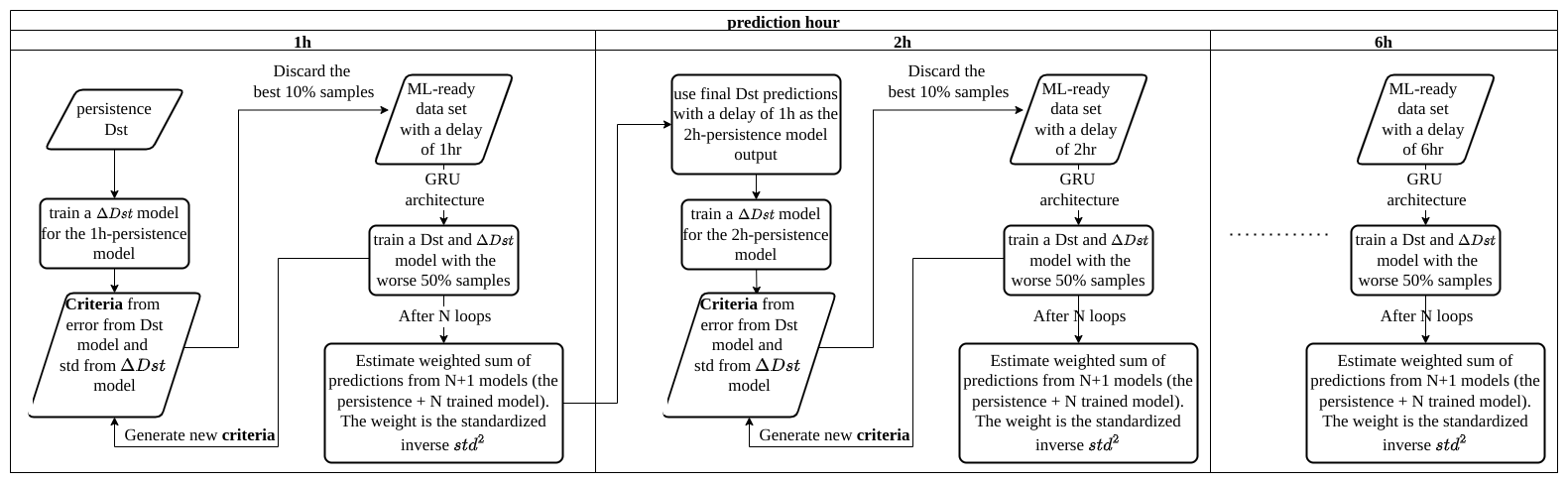}
  \caption{Workflow of the multi-fidelity boosting method. It should be noted that, in order to mitigate overfitting, the best 10\% percentage samples will be discarded before each training.}
  \label{fig:boost_method}
\end{figure}

\begin{table}[!htbp]
  \centering
  \caption{RMSE of the developed model with different boost criteria.}
  \begin{tabular}{c | c | c c c c}
    \hline
    Predict  & Boost  &  & RMSE(nT) & & \\
     time& criteria &  all & ($-\infty$, -100] & (-100, -50] & (-50, $\infty$)\\
       \hline
      & error & 2.72 & 5.02 & 3.80 & 2.07 \\
      1h& $\sigma$ & 2.74 & 5.78 & 3.27 & 2.06 \\
      & z-score & \textbf{2.59} & \textbf{4.84} & \textbf{3.31} & \textbf{2.00} \\
      \hline
      & error & 5.46 & 10.63 & 7.80 & 4.12 \\
      2h& $\sigma$ & 5.98 & 12.68 & 7.09 & 4.69 \\
      & z-score & \textbf{3.81} & \textbf{7.57} & \textbf{3.89} & \textbf{3.13} \\
      \hline
      & error & 10.19 & \textbf{20.83} & 11.34 & 8.14 \\
      3h& $\sigma$ & 10.39 & 22.10 & 10.88 & 8.24 \\
      & z-score & \textbf{8.22} & \textbf{16.99} & \textbf{10.37} & \textbf{6.00} \\
      \hline
      & error & 12.04 & 24.63 & 13.78 & 9.43 \\
      4h& $\sigma$ & 12.68 & 26.37 & 13.28 & 10.01 \\
      & z-score & \textbf{10.44} & \textbf{22.19} & \textbf{11.85} & \textbf{7.99} \\
  \end{tabular}
  \label{tab:boost}
\end{table}

In order to precisely assess the accuracy of a model, it is important that the performance metrics are computed on a test set independent from the training set, so-called unseen data. Hence, making sure that the machine learning algorithm does actually learn meaningful patterns and does not merely memorize the training data. A `Leave one out' technique is adopted here. That is a K-fold cross validation taken to its logical extreme, with K equal to N, the number of selected storm cases. That means that the proposed model is trained on all the data except for one storm window and a prediction is made for that left-out storm. The procedure is repeated N times. Finally, the metrics are computed as averages over the N models. In this study, each of the 67 storm windows reported in Table \ref{tab:storms} constitutes a fold.  Root-mean-square error (RMSE) is used as the main metric to assess the accuracy of the proposed model. The continuous $Dst$ prediction can also be transformed to a  binary label upon defining a threshold, i.e. -100 nT in this study. In this way we can use standard metrics for binary classification such as the True Skill Statistic (TSS) and Matthews correlation coefficient (MCC)\cite{camporeale2020gray}. The MCC score is a reliable statistical rate that produces a high score only if the prediction obtained good results in all of the four confusion matrix categories --- true positive (TP), False positive (FP), True Negative (TN) and False Negative (FN) --- proportionally both to the size of positive elements and the size of negative elements in the data set \cite{baldi2000assessing}. TSS is a useful metric that combines both types of information and should be as close as possible to 1. Those metrics have shown some advantages over the F1 score and accuracy in binary classification evaluation \cite{chicco2020advantages}.


\section{Results}
\label{sec:results}

In this section, the proposed boosting model is first compared against a simple GRU model. 

\begin{figure}[!htbp]
    \centering
    \includegraphics[width=0.48\textwidth]{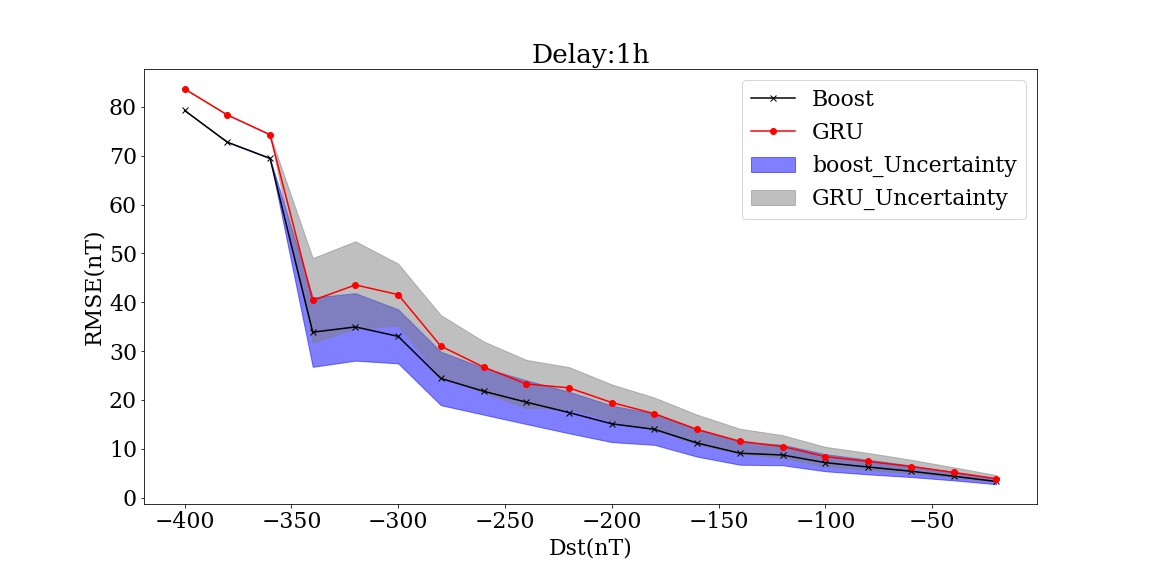}
    \includegraphics[width=0.48\textwidth]{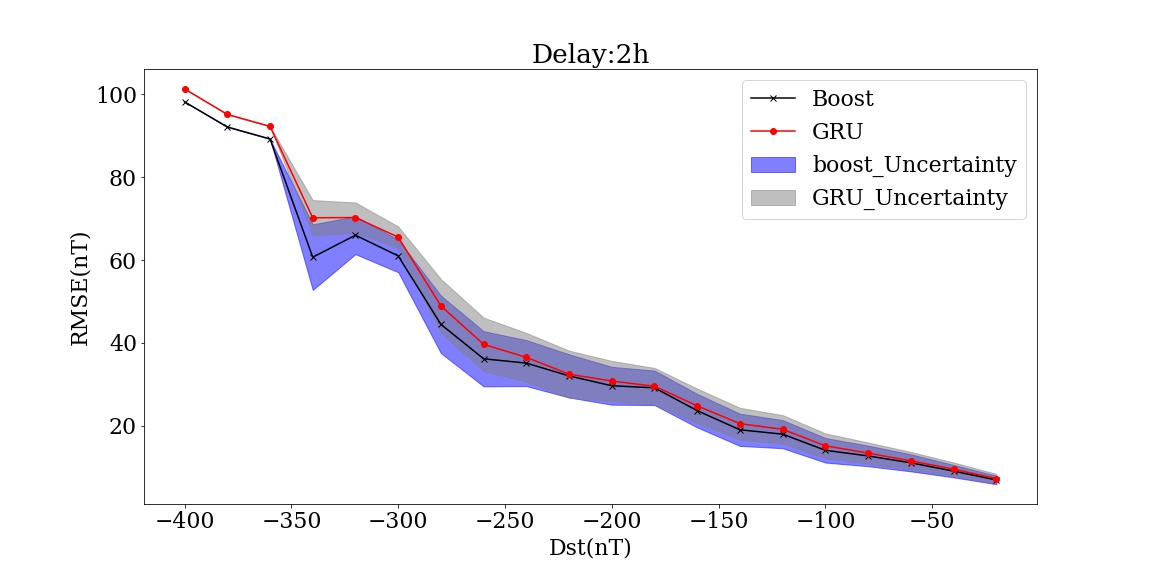}
    \includegraphics[width=0.48\textwidth]{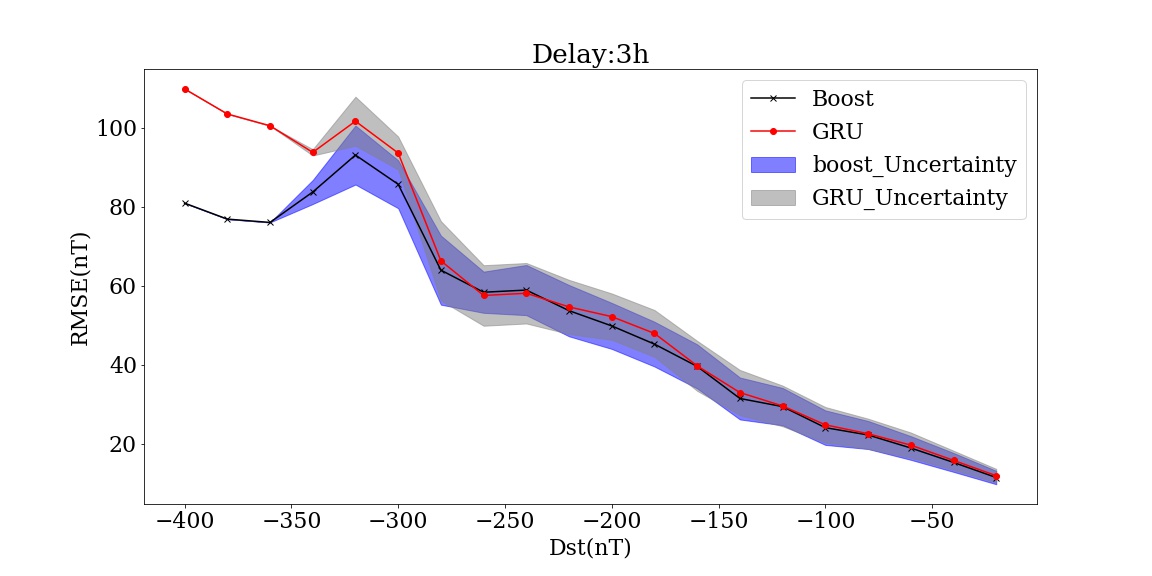}
    \includegraphics[width=0.48\textwidth]{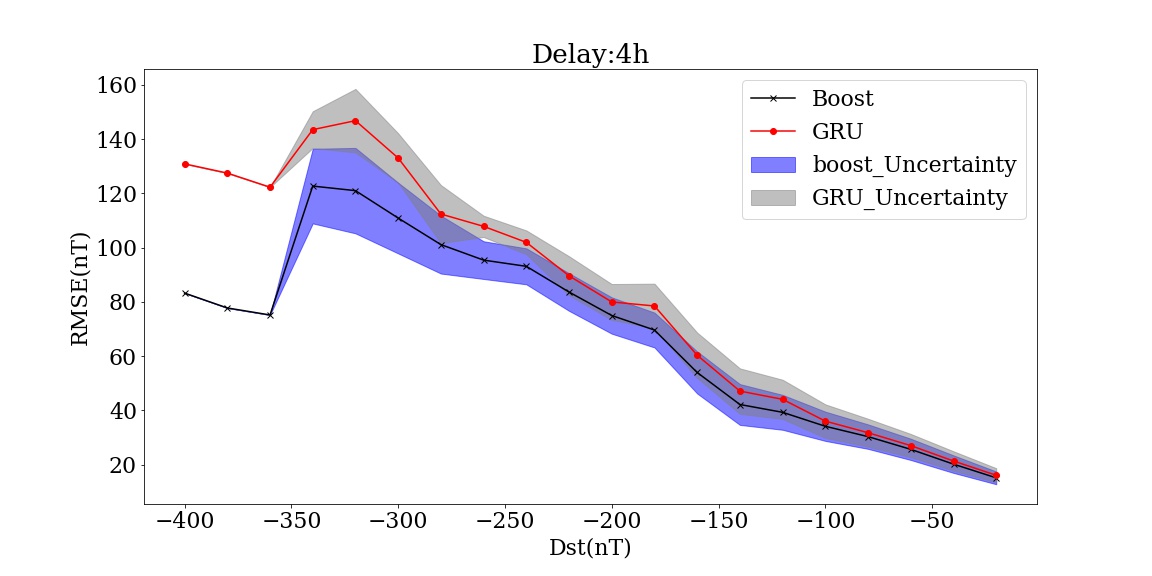}
    \includegraphics[width=0.48\textwidth]{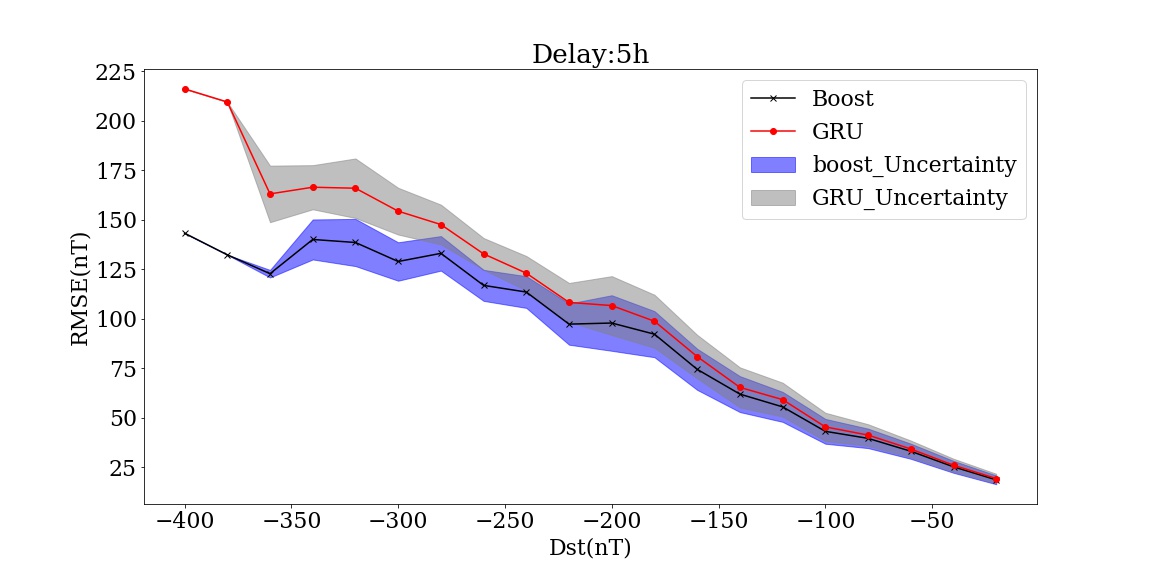}
    \includegraphics[width=0.48\textwidth]{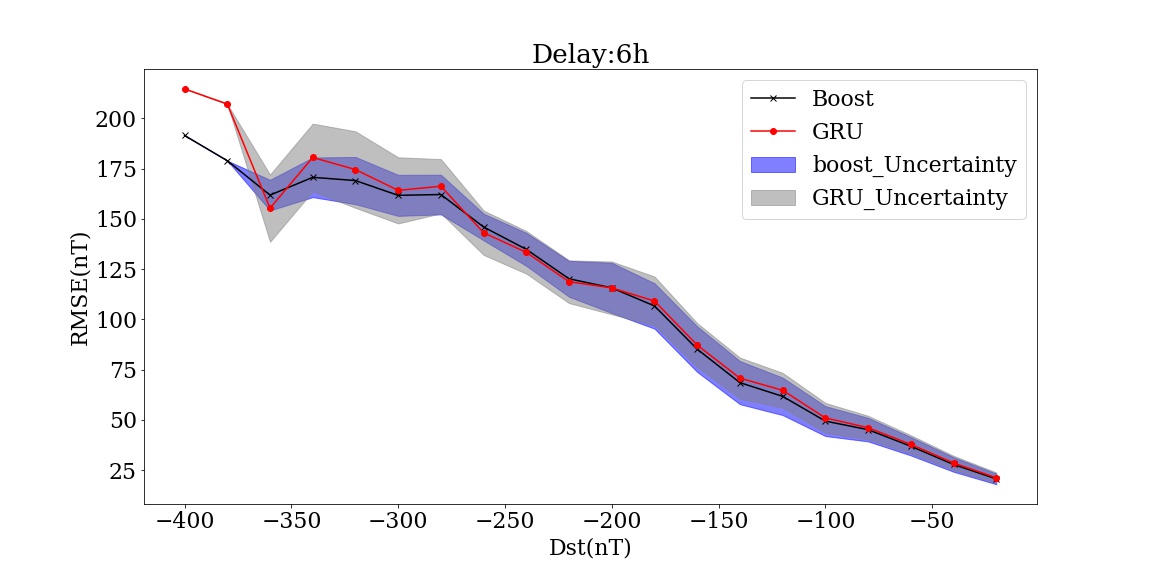}
    \caption{RMSE of the proposed model with 1-6 hrs time delay. For each panel, the horizontal axis is the $Dst$ threshold, and the vertical  axis is the RMSE of all samples below that threshold, evaluated in the test set. In each panel, black line denotes average boost results, red lines shows average GRU results, and purple and grey bars are the uncertainty for boosting and GRU results, respectively.}
    \label{fig:SH_Resi}
\end{figure}

Fig. \ref{fig:SH_Resi} shows a statistic analysis of the proposed model with 1-6 hrs time delay. In each panel, each dot represents the mean RMSE computed over all test samples with $Dst$ smaller than a given $Dst$ threshold (on horizontal axis). The red line shows the average RMSE of GRU results, and the grey bar is the corresponding RMSE standard deviation. The red line and the purple bar show the corresponding average RMSE and standard deviation for the developed boosting method, respectively. 

\begin{figure}[!htbp]
  \centering
  \includegraphics[width=\textwidth]{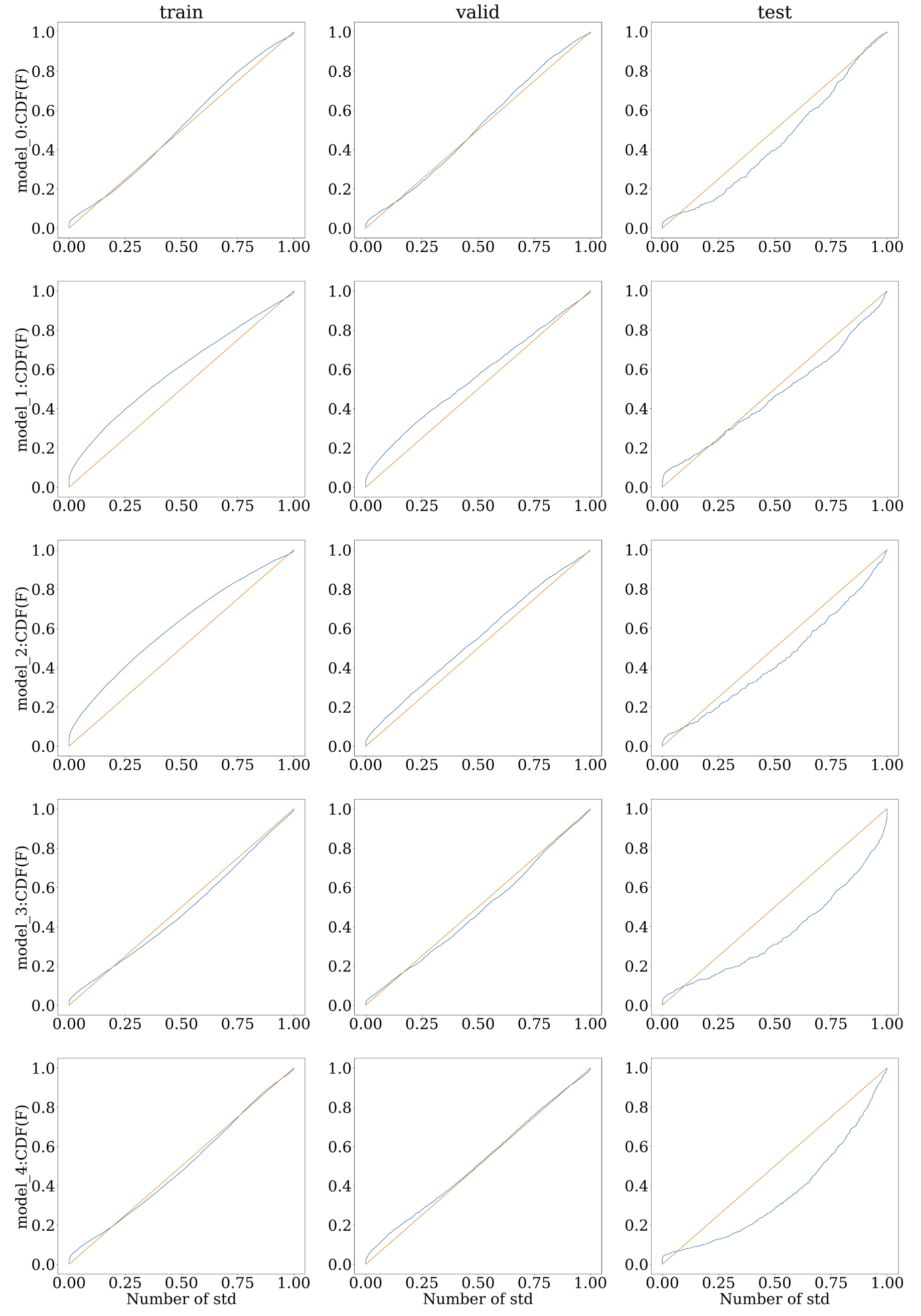}
  \caption{Quantile-quantile (Q-Q) plots (in blue) averaged over all storm events (6-hr ahead predictions for the final multi-fidelity results). From left to right, panels are QQ-plots for training, validation and test set. In each panel, the Q-Q line is blue, and the orange line is a 45-degree reference line (i.e., y=x). }
  \label{fig:QQ}
\end{figure}

It is clear that the boosting method significantly outperforms a single GRU method, for all delays. The corresponding Q-Q plots are shown in Fig. \ref{fig:QQ}. If z-scores estimated by the ACCRUE method are distributed as a perfect Gaussian, then the Q-Q lines (in blue) would perfectly overlap the diagonal line (in orange). In Figure \ref{fig:QQ}, the Q-Q lines well agree with the diagonal lines in the training and validation sets. This correlation is slightly worse in the test set because the distribution of the test set might differ from that of the training and validation sets.

\subsection{Storm Case Study}
\label{subsec:storm_case}

In this section we would like to investigate two typical storm cases. 1) The 2003-Halloween storm, the biggest storm in the past 20 years, from 2003-11-07 to 2003-11-29; and 2) The 2021-Halloween storm, a recent storm with Dst peak $< -100$ nT, between 2021-11-02 to 2021-11-08.

\subsubsection{2003-Halloween storm}
\label{subsub:2003-Halloween}

\begin{figure}[!htbp]
    \centering
    \includegraphics[width=\textwidth]{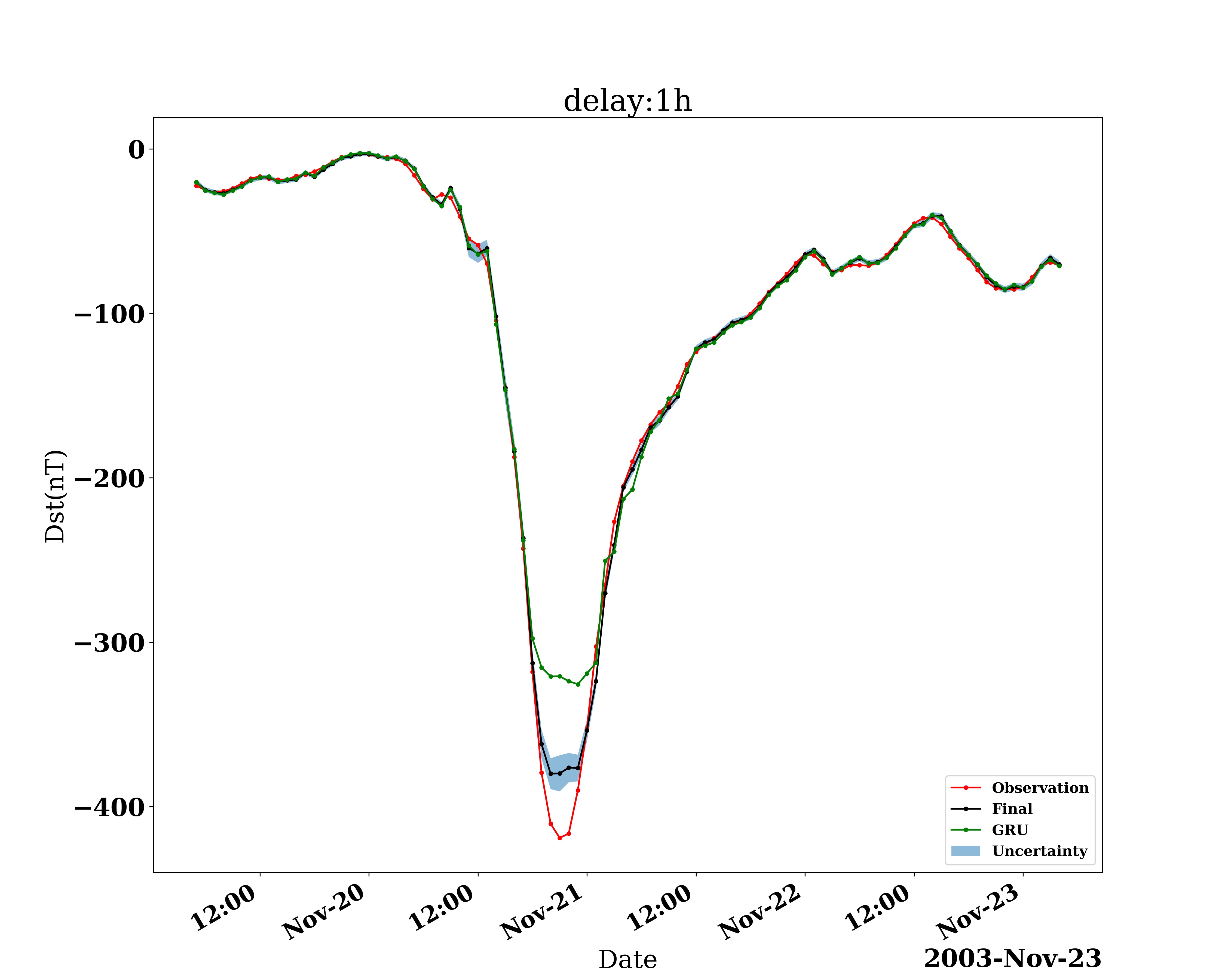}
    \includegraphics[width=\textwidth]{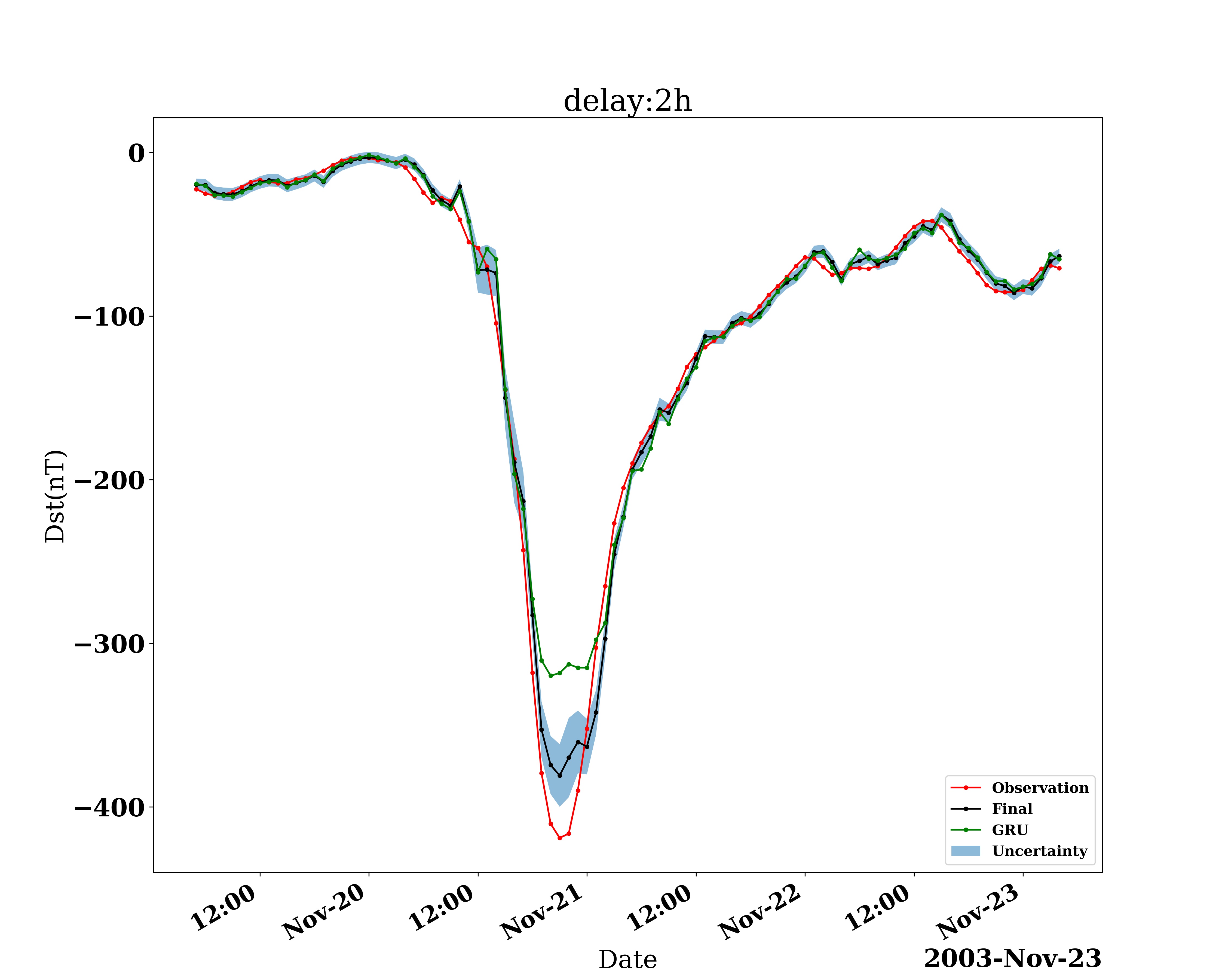}
    \caption{$Dst$ predictions between 2003-11-09 and 2003-11-29 with 1-to-2-hr delays. In each panel, red line is the observed $Dst$. Green and black lines are GRU and multi-fidelity predictions. Blue bars are the multi-fidelity uncertainty.}
    \label{fig:2003-Halloween-1}
\end{figure}

\begin{figure}[!htbp]
    \centering
    \includegraphics[width=\textwidth]{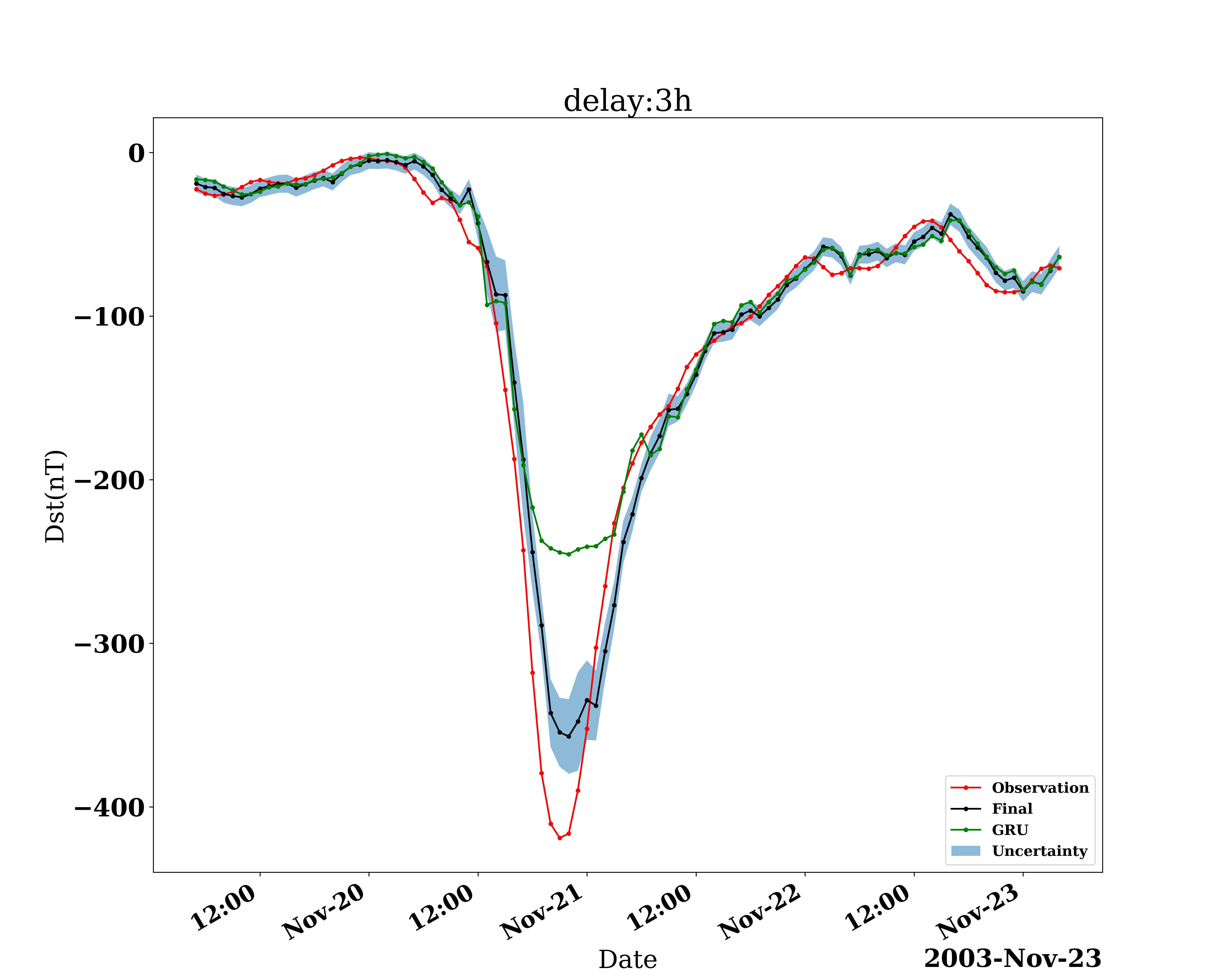}
    \includegraphics[width=\textwidth]{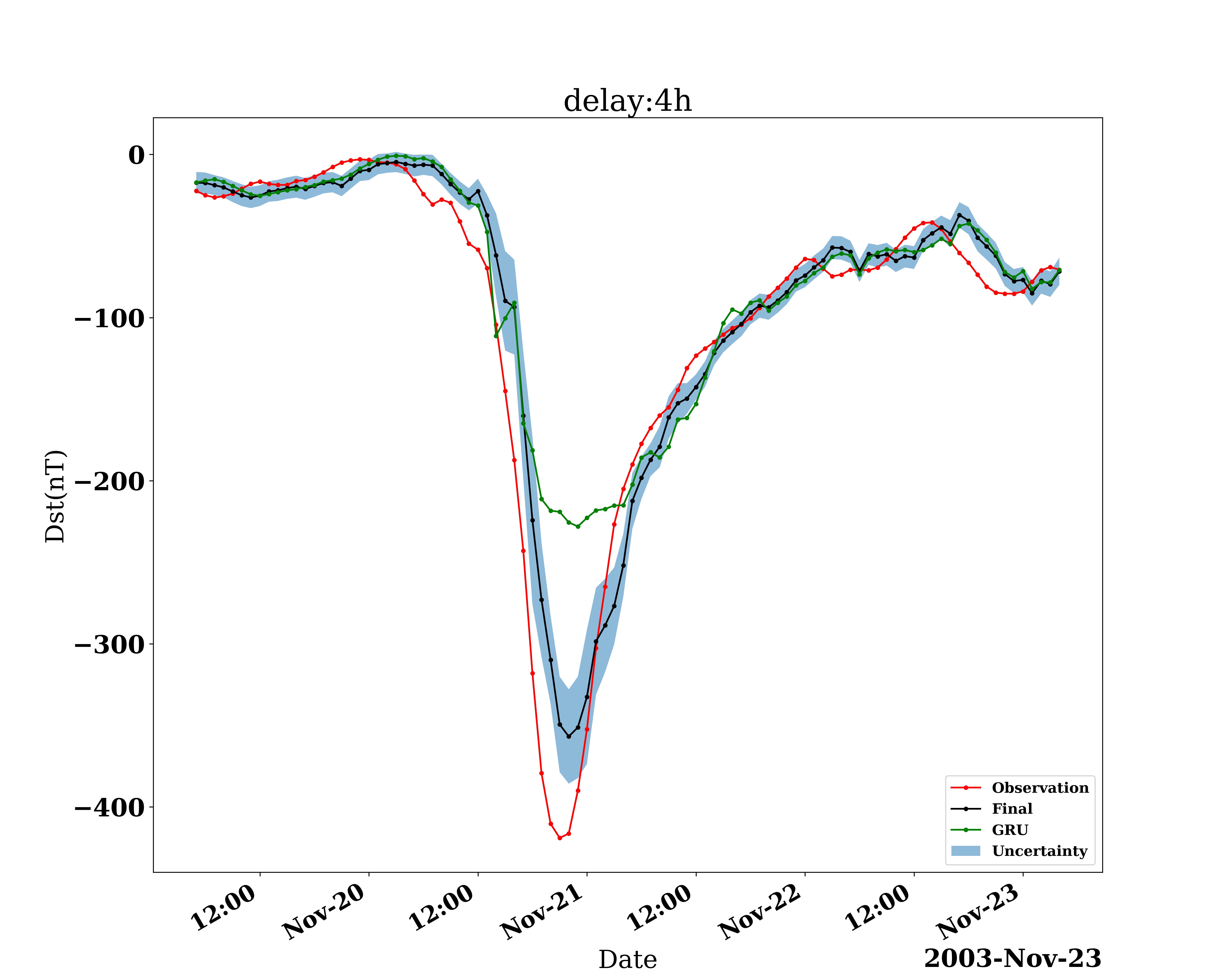}
    \caption{Similar to Fig. \ref{fig:2003-Halloween-1}, except for 3-to-4 hrs}
    \label{fig:2003-Halloween-2}
\end{figure}

\begin{figure}[!htbp]
    \centering
    \includegraphics[width=\textwidth]{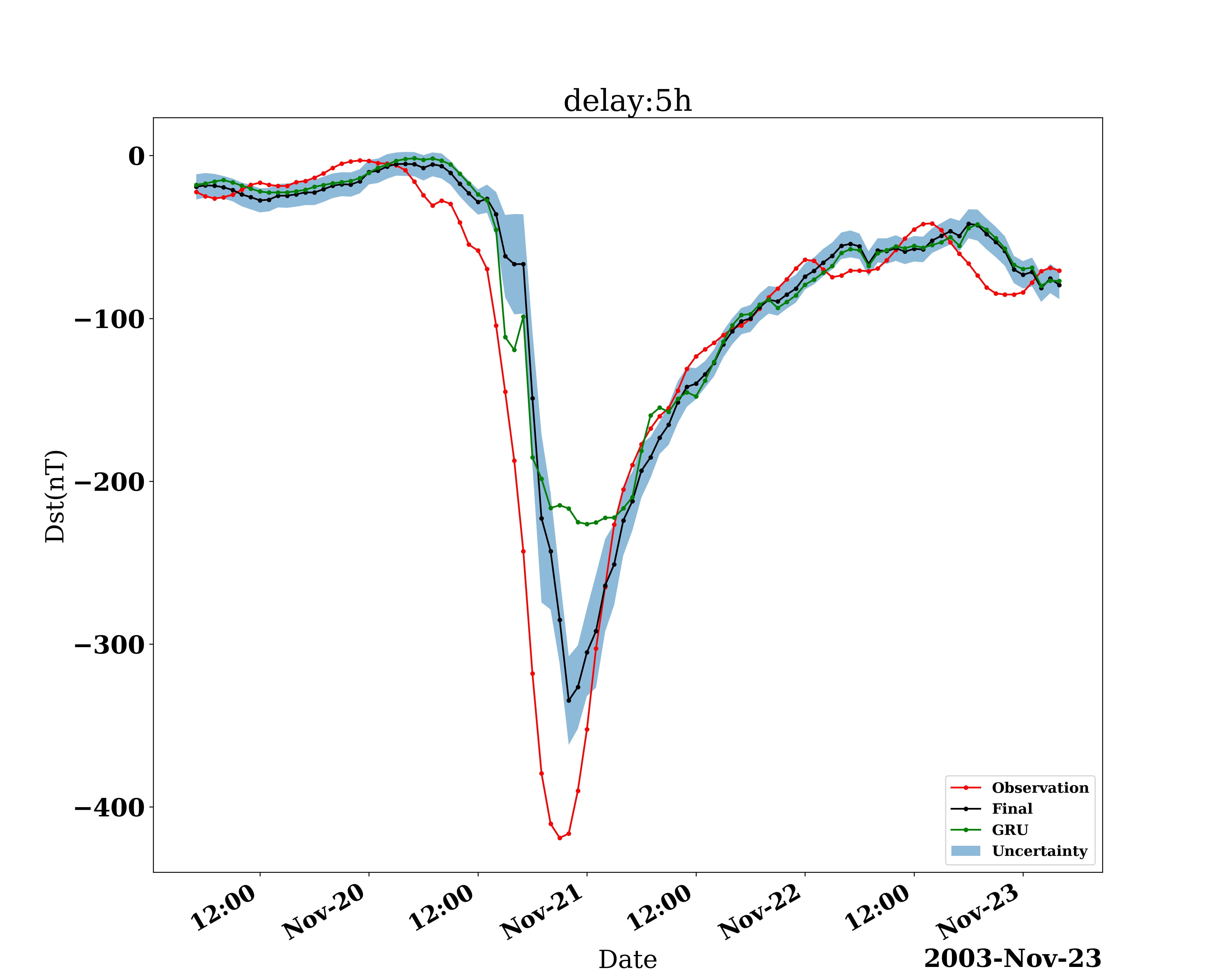}
    \includegraphics[width=\textwidth]{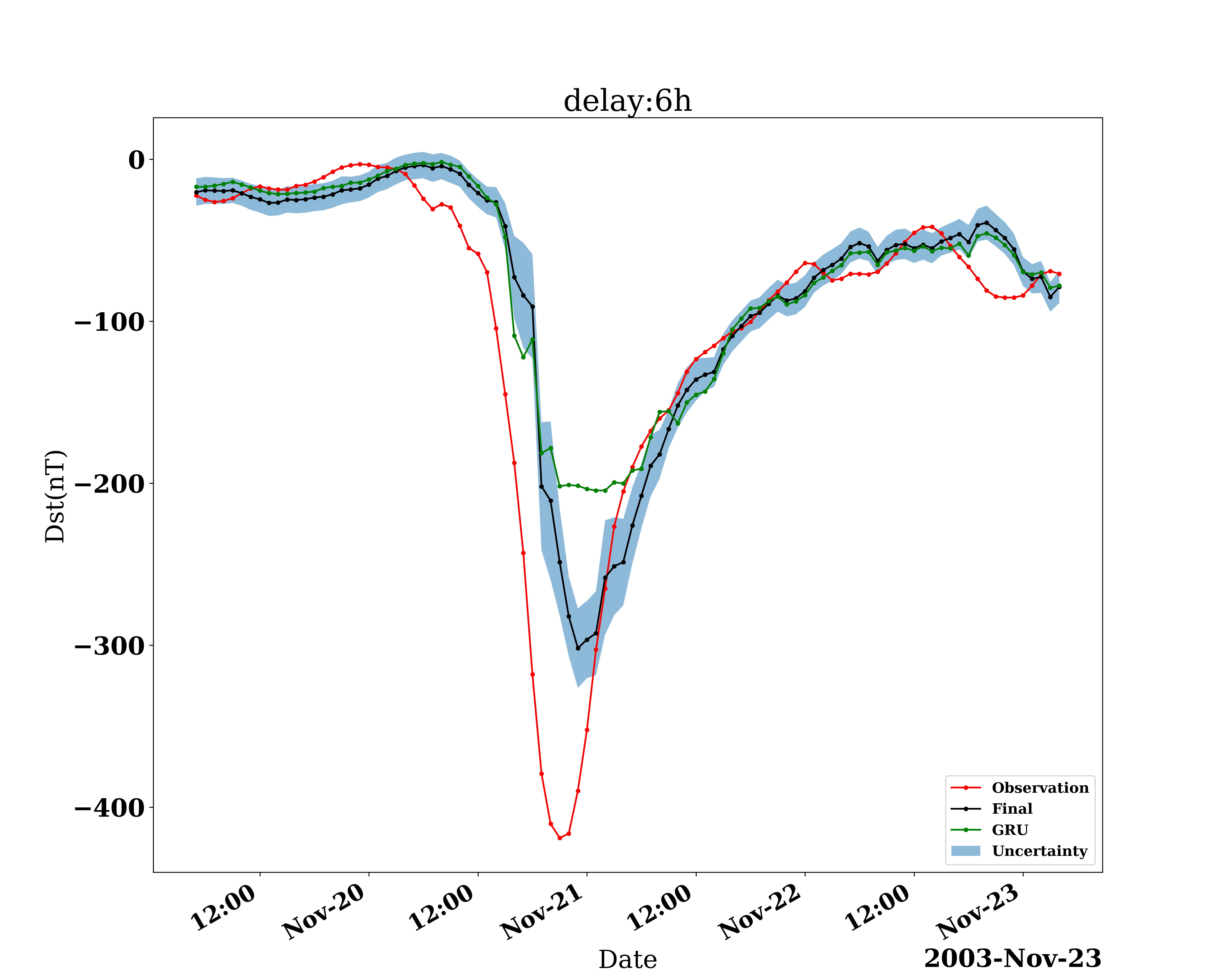}
    \caption{Similar to Fig. \ref{fig:2003-Halloween-1}, except for 5-to-6 hrs}
    \label{fig:2003-Halloween-3}
\end{figure}

Figure \ref{fig:2003-Halloween-1} \& \ref{fig:2003-Halloween-2} \& \ref{fig:2003-Halloween-3} display the $Dst$ predictions of the proposed model (red for GRU and black for multi-fidelity boosting results) and the corresponding observed $Dst$ (green) during the 2003-Halloween storm. The time delay of each panel is corresponding to the same panels in Fig. \ref{fig:SH_Resi}. 

The GRU model can better predict the shift time, especially between the main phase and recovery phase. Meanwhile, the peak $Dst$ is always better predicted by the persistence model. This proves that the multi-fidelity boosting model can take advantage of both models, and further verifies our findings presented in Fig. \ref{fig:SH_Resi}. Similar plots for all the other storm events used in this study are included as supplementary information.  

\subsubsection{2021-Halloween storm}
\label{subsub:2021-Halloween}

\begin{figure}[!htbp]
    \centering
    \includegraphics[width=\textwidth]{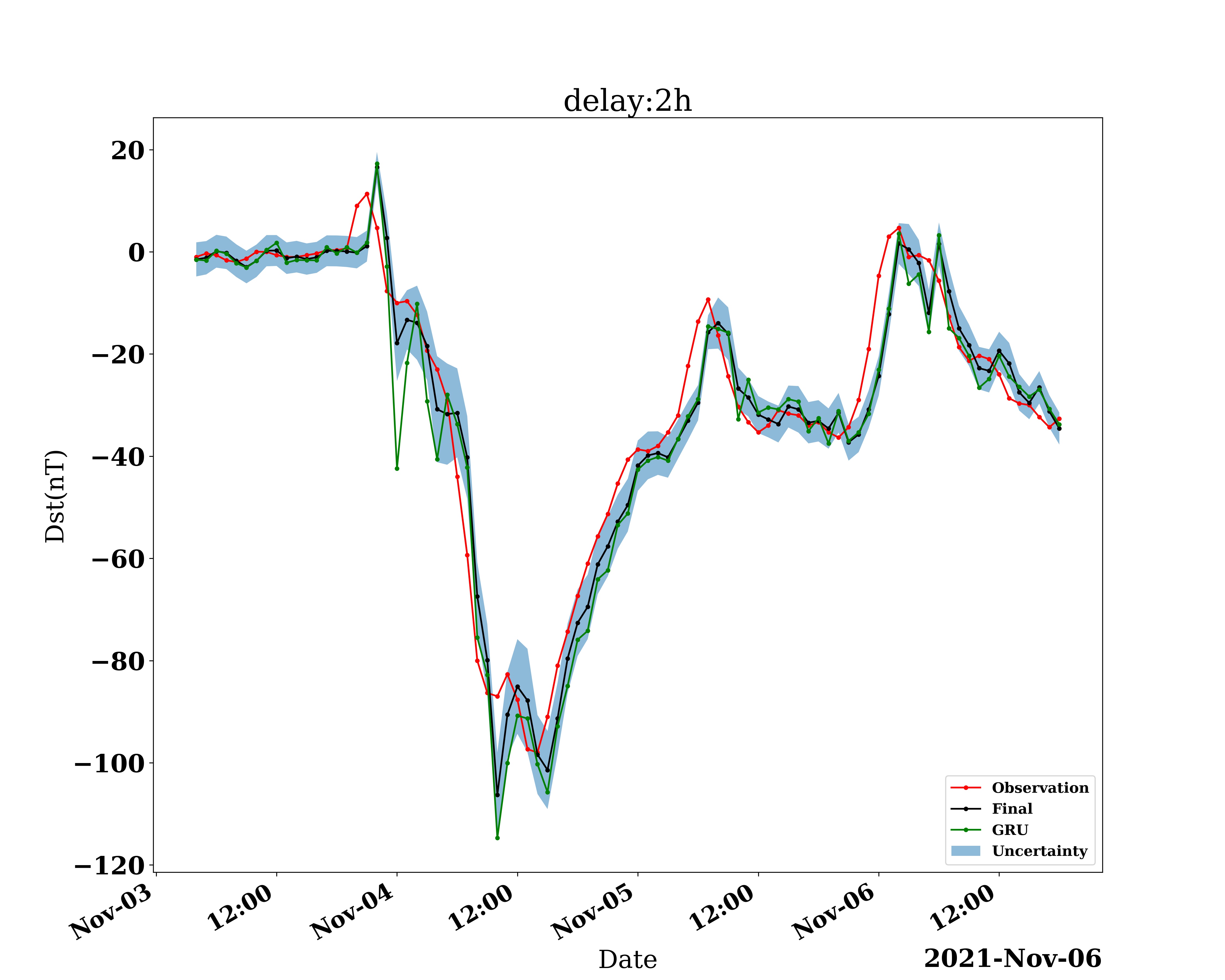}
    \includegraphics[width=\textwidth]{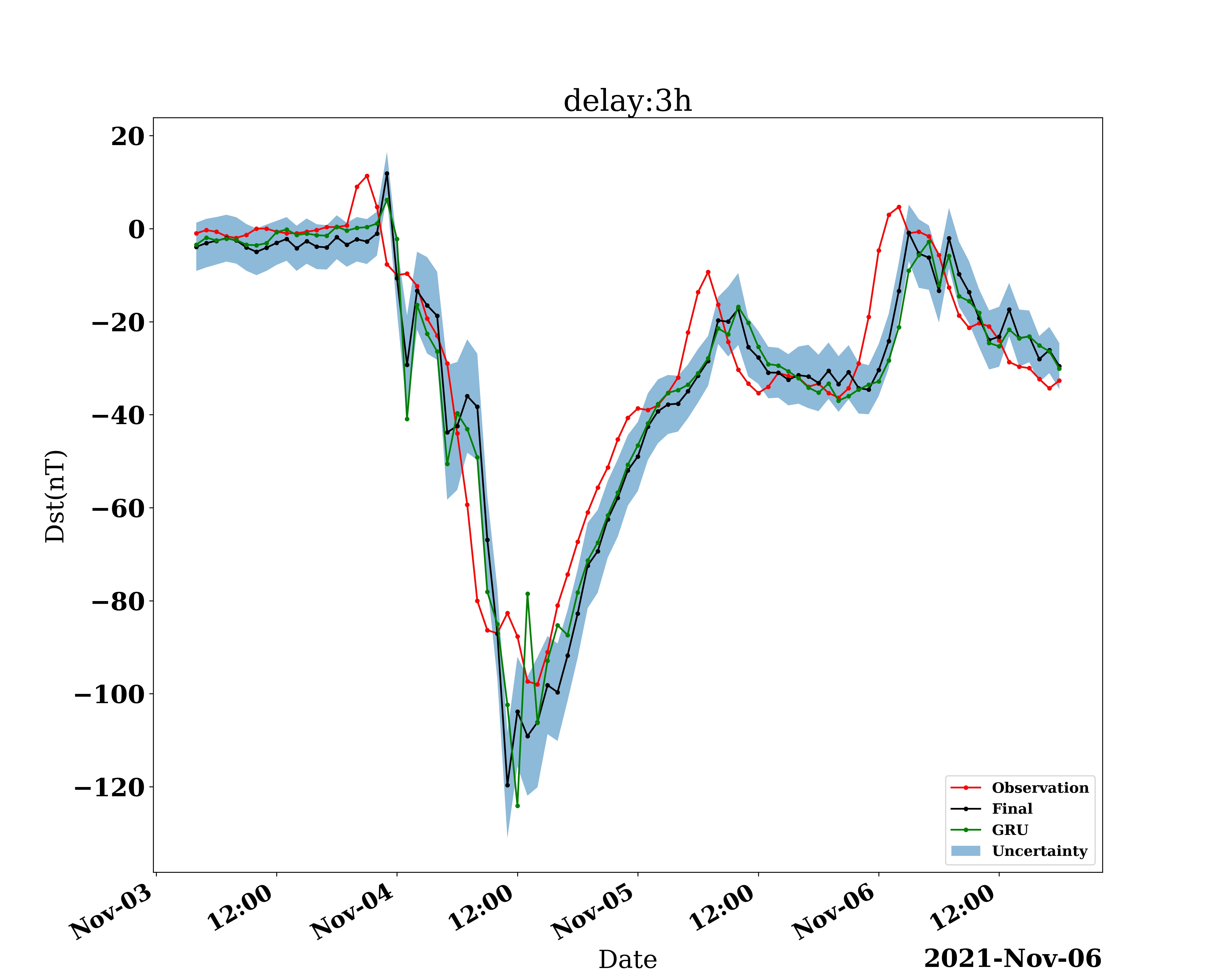}
    \caption{Similar to Fig. \ref{fig:2003-Halloween-1}, except 2-to-3 hrs ahead $Dst$ predictions around 2021 Halloween.}
    \label{fig:2021-Halloween}
\end{figure}

The same figures are plotted in Figure \ref{fig:2021-Halloween}. We could see that the proposed model can also better predict the $Dst$ during the main phase of this 2021-Halloween storm than the persistence model. It should be noted that the test set is ACE measurements but the model is trained using OmniWeb data. Hence, it is possible that the performance can be further improved by using ACE measurements to train the model for real forecast. These results will be exhibited on our group website soon.

\section{Summary \& Outlook}
\label{sec:summary-outlook}

We have developed a multi-fidelity boosting model to predict $Dst$ during geomagnetic storm periods, 1-6 hrs ahead. 
Sixty-seven selected storm events were chosen during a long‐span historical data set ($\sim$ 20 years), between 2000-01-01 and 2020-01-01. One of the crucial points of this work is that an innovative multi-fidelity boosting method is developed to enhance a simpler prediction algorithm (here based on GRU networks), especially during those `super' storms as defined in Sec. \ref{sec:D&M}. The prediction might be further improved by considering the new variables, such as $F10.7$ and $\theta_c$ for super periods. We have shown that this proposed model provides a good RMSE ($8.22 nT$) for predicting $Dst$ up to 3 hrs ahead during strong storm periods. 
The prediction becomes worse with a longer lead time (13.54 nT at 6 hrs) because of the increasing errors of the baseline model. We have also discussed how this model performs during the 2003/2017 Halloween storms. The $Dst$ peak can be well captured by the developed model. 


\acknowledgments

This project has been developed in the framework of the National Aeronautics and Space Administration under grants 80NSSC20K1580 and  80NSSC20K1275. We thank OMNIWeb for providing the $Dst$ data (https://omniweb.gsfc.nasa.gov/) and SuperMAG for ground magnetometer data(https://supermag.jhuapl.edu/). All the results and codes have been made available on https://github.com/ML-Space-Weather/LiveDst, an active DOI will be provided if this manuscript is accepted.
\bibliography{Ref}

\newpage

\end{document}